# AI as a teaching tool and learning partner

S Watterson[1*], S Atkinson[2], E Murray[1], A McDowell[2]

[1]School of Medicine, Ulster University, Northlands Road, Derry~Londonderry, BT48 7JL, UK

[2]School of Biomedical Science, Ulster University, Cromore Road, Coleraine, BT52 1SA, UK

*s.watterson@ulster.ac.uk

## Abstract

The arrival of AI tools and in particular Large Language Models (LLMs) has had a transformative impact on teaching and learning and institutes are still trying to determine how to integrate LLMs into education in constructive ways. Here, we explore the adoption of LLM-based tools into two teaching programmes, one undergraduate and one postgraduate. We provided to our classes (1) a LLM-powered chatbot that had access to course materials by RAG and (2) AI-generated audio-only podcasts for each week's teaching material. At the end of the semester, we surveyed the classes to gauge attitudes towards these tools. The classes were small and from biological courses. The students felt positive about AI generally and that AI tools made a positive impact on teaching. Students found the LLM-powered chatbot easy and enjoyable to use and felt that it enhanced their learning. The podcasts were less popular and only a small proportion of the class listened weekly. The class as a whole was indifferent to whether the podcasts should be used more widely across courses, but those who listened enjoyed them and were in favour.

## Introduction

There has been an explosion in the availability of AI tools with the potential to impact on teaching and learning. Large Language Models (LLMs) have facilitated natural language interactions in a way that simplifies how users interact with swathes of data and knowledge. The most well-known AI tools based on LLMs include ChatGPT, CoPilot, Gemini, Claude and Perplexity.AI, and these services generate text responses statistically based on large volumes of training text. They can be enhanced with new text, often from concurrent web searches, to improve the relevance and factual accuracy of the responses.

Much of the narrative around AI tools has been negative, focussing on plagiarism in assessment. We wanted to explore their positive potential and evaluate how AI tools could be used as teaching tools and learning partners by students in the Life Sciences, in particular determining how comfortable students felt working with AI tools.

## Context and Background

LLMs have recently been suggested as tutors and learning partners (Grassucci et al., 2025) and the supplementation of LLMs with teaching materials has been considered (Li et al., 2025; Shojaei et al., 2025). Bespoke mechanisms for soliciting feedback on their impact on teaching have been proposed (Fuligni et al., 2025). We aimed to introduce an AI powered chatbot and a series of AI-powered podcasts into BIO123 (BSc Personalised Medicine) and BIO711 (MSc Biotechnology) to support students in their learning. At the end of the semester, we sought feedback from the class to evaluate how they felt about using the AI-chatbot and podcasts and whether they felt any benefit. To implement this study, we enhanced a LLM with module information and learning materials, including the module handbook, lecture notes, supplementary reading, assessment guidelines (but not assessments themselves) and rubrics. The LLM was made available via an AI-powered chatbot to class members who were able to ask it questions about anything related to the module content and organisation. The chatbot could provide alternative explanations, revision timetables, flashcards, etc. Example use-cases were provided.

Module materials were provided to the LLM by Retrieval Augmented Generation, a process that ensured they were not incorporated into the underlying model and supported factual recall. We updated the LLM/chatbot weekly so that its knowledge base grew with the class. AI-generated podcasts were provided that contained a discussion of each week's materials. Both reused module materials, so students who chose not to participate were not disadvantaged.

## Key Insights or Strategies

*Students are embracing AI tools*

From a cohort of 41 students, we received 28 responses, though some were incomplete. Agreement with the statement that "I feel that AI tools will make a positive or negative contribution to society" with 1=very negative, 5=very positive, scored a mean of 4.0 (21 responses, Fig. 1A). In an agree/disagree response, 75% agreed that "I do regularly use any AI tools outside of my courses" (16 responses, Fig. 1B). Students felt that they benefitted from using these tools. 100% agreed that "I do feel that I gained from using the AI-chatbots and AI-generated podcasts" (14 responses, Fig. 1C).

However, students are not confident. When asked how they would rate their skills with AI tools (1=poor and 5=high), the mean score was 3.2 (21 responses, Fig. 1D).

***Students felt positive about the AI-powered chatbot***

Their chatbot experience was positive. Agreement with "I found it easy to understand how to use the AI-chatbot", scored a mean of 4.1 (24 responses, Fig. 1E). Agreement with "I enjoyed using the AI-chatbot", scored a mean of 4.2 (24 responses, Fig. 1F).

100% of the students who responded agreed that "Using the AI-chatbot enhanced my knowledge of the module learning content" in an agree/disagree response (20 responses, Fig. 1G).

***AI-generated podcasts were less popular***

The majority of respondents used the podcasts occasionally or never. Few students listened weekly (25 responses, Fig. 1H). However, amongst those who used the podcasts, opinion was positive. For respondents who listened at least occasionally, agreement with the statement "I enjoyed listening to the AI-generated podcast" scored a mean of 4.3 (15 responses, Fig. 1I).

Whether students would like to see more podcasts was mixed. Amongst all responses, agreement with "I would like to see AI-generated podcasts made available in more of my modules" scored a mean of 3.5 (22 responses, not shown). However, the students who listened at least occasionally, agreed with a mean score of 4.1 (15 responses, not shown).

## Challenges and Solutions

The specific AI-chatbot used was powered by Google and required a (free) Google account to gain access. Students may have been reluctant to register or share their Google account details to gain access and this may explain why not all the class used the AI-chatbot. However, Google is currently the only provider of tools with which we can generate the chatbot without coding it directly, making it an efficient solution. The AI-generated podcasts had no such restrictions and were provided on Blackboard Learn as audio files in .wav format. Whilst they could be listened to in a browser, a more natural setting for listening to podcasts is while exercising or commuting and perhaps the steps involved in downloading and transferring the podcasts to a phone or audio device inhibited their wider consumption.

The modules were small, but chosen due to our roles as module coordinators. They came from scientifically advanced Life Science courses with cohorts we would ordinarily expect to feel positive about technology. Whilst these cohorts may be representative of science and technology courses, they may not be representative of the wider student body.

Whilst this is a limited study, questions about how we integrate AI tools into teaching and learning must be addressed and this exercise has provided us with preliminary data about student attitudes and engagement.

## Reflections and Future Considerations

AI is exploding into our working lives and our graduates will need to be equipped to use it as they move into the workforce. At the same time, the potential they offer to improve teaching and learning, and to increase efficiency in its delivery, will be beneficial to HE staff. However, presently we are still navigating how best to use them.

These tools have been introduced to open up new ways to access knowledge but, paradoxically, without adequate support the benefit they provide to HE will be strongly biased towards technologically literate students and modules taught by technologically literate staff. Barriers to their adoption for staff include a lack of understanding of how AI can be harnessed, limited case studies of its use in education and technological barriers to its implementation. These barriers are also likely to be felt by some of our students.

AI tools will allow students to interact in new ways with module material and help with module organisation. They will act as a personal tutor and revision partner, generate alternative explanations of lecture content, create revision materials and revision schedules, and generate multimodal learning resources. As a result, AI will undoubtedly become embedded in our future teaching and learning infrastructure. However, questions remain. How do we teach practical skills to students in the use of AI-tools? How do we upskill staff? How do we adapt assessments to ensure that access to AI doesn't impinge on evaluating understanding?

## Conclusion

Our conclusions from these limited cohorts suggest that there is an appetite amongst students to use AI tools to support their learning and that they are quite capable of using AI tools when they are provided.

**Key takeaways**

- Students found our AI powered module-specific chatbots easy and enjoyable to use and unanimously reported that they enhanced their learning.
- Whilst the AI-powered podcasts were not as popular, they were also valued as an extra modality in their learning when adopted.
- However, a lack of confidence in AI skills suggests that some of the class may need to be supported in learning how best to use these tools.

**Author Bios**

Dr Steven Watterson is a lecturer in computational biology in the School of Medicine with interests in Systems Biology, Machine Learning and AI, especially with application to Cardiovascular Disease.
https://www.linkedin.com/in/stevenwatterson/
s.watterson@ulster.ac.uk

Dr Andrew McDowell is a Senior Lecturer in Molecular Microbiology in the School of Biomedical Sciences. His research is focused on the role of the resident human microbiota in health and disease. He also has an interest in the use of technology to enhance student learning and outcomes.
https://www.linkedin.com/in/dr-andrew-mcdowell-a53511224/
a.mcdowell@ulster.ac.uk

Dr Elaine Murray is a Senior Lecturer in Personalised Medicine (Mental Health) in the School of Medicine at Ulster University with interests in biomarker discovery and novel AI applications particularly in neuroscience and mental health.
https://www.linkedin.com/in/elaine-murray-4778b210/
e.murray@ulster.ac.uk

Dr Sarah Atkinson is a Senior Lecturer and Academic Lead for Education in the School of Biomedical Science at Ulster University; her interests include AI in education.
https://www.linkedin.com/in/sarah-atkinson-23441a17/
s.atkinson@ulster.ac.uk

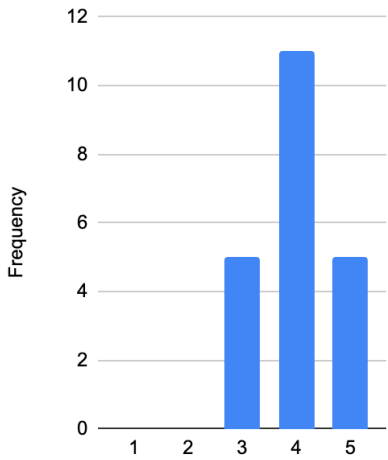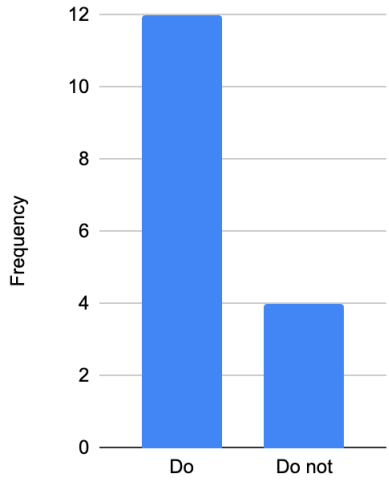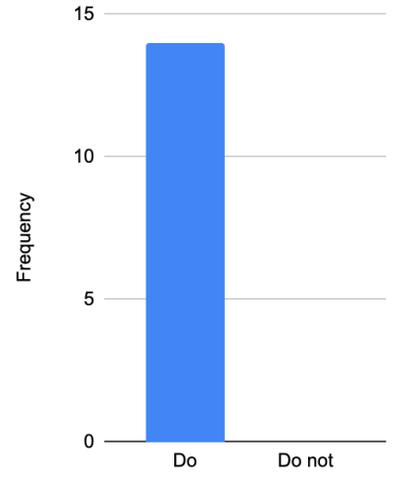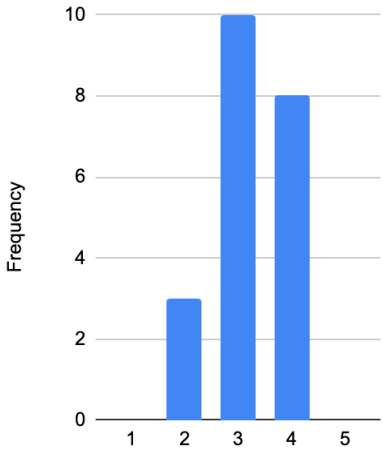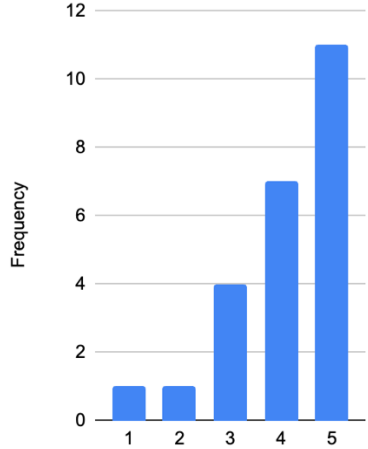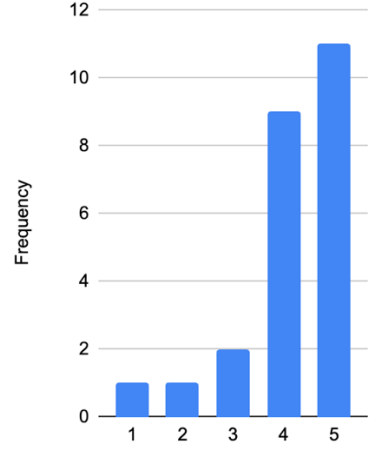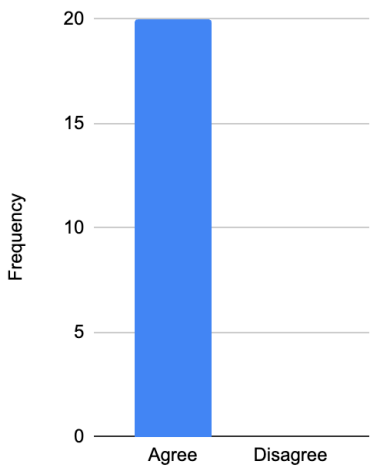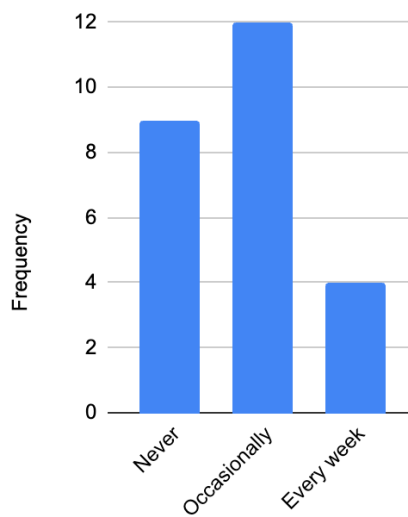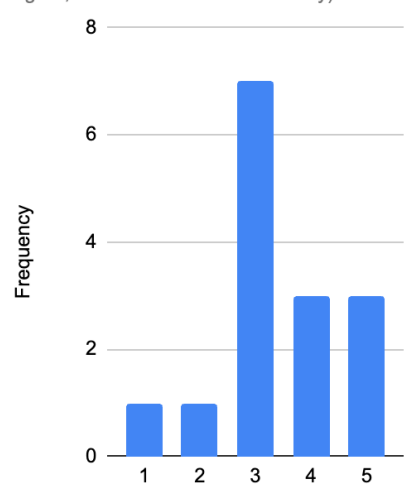

Figure 1. Survey results describing student views on the AI tools introduced.